\documentclass[]{jfm_SI}

\usepackage{newtxtext}
\usepackage{newtxmath}
\usepackage{natbib}
\usepackage{hyperref}
\usepackage{epstopdf, epsfig}
\usepackage{comment}
\usepackage{pdflscape}
\usepackage{tabularx}
\usepackage{lipsum}
\usepackage{amssymb}
\usepackage{amsmath}
\usepackage{overpic}
\usepackage{epstopdf}
\usepackage[T1]{fontenc}
\usepackage[utf8]{inputenc}
\usepackage{color,soul}
\usepackage{placeins}
\usepackage{hyperref}
\usepackage{caption}
\usepackage{booktabs}
\usepackage{makecell}

\usepackage{stfloats}

\usepackage{graphicx}
\usepackage{newtxtext}
\usepackage{newtxmath}
\usepackage{natbib}
\usepackage{hyperref}
\hypersetup{
    colorlinks = true,
    urlcolor   = blue,
    citecolor  = black,
}

\newcommand{\RomanNumeralCaps}[1]

\title{Semi-analytical eddy-viscosity and backscattering closures for 2D geophysical turbulence}

\author{Yifei Guan\aff{1,2}\corresp{\email{guany@union.edu}},
  Pedram Hassanzadeh\aff{1}}
  
\affiliation{\aff{1} University of Chicago, Chicago, IL 60637, USA 
\aff{2} Union College, Schenectady, NY 12308, USA}

\begin{document}
\maketitle

\begin{abstract}
Physics-based eddy-viscosity and backscattering closures are widely used for large-eddy simulation (LES) of geophysical turbulence, but their key parameters are often chosen empirically. Here, we develop a semi-analytical framework for estimating these parameters in 2D geophysical turbulence. Specifically, we extend a Lilly-type scaling argument, previously used for 3D turbulence, to 2D geophysical turbulence and obtain closed-form estimates, up to an amplitude constant, for the coefficients of the Leith and Smagorinsky eddy-viscosity closures, a biharmonic eddy-viscosity closure, and the Jansen--Held backscattering closure with a prescribed backscattering fraction. The amplitude constant appears in the turbulent kinetic energy direct-cascade spectrum and can be diagnosed from a few direct numerical simulation (DNS) or eddy-resolving snapshots. For the $\beta$-free cases, the diagnosed amplitude constant is consistent with previous theoretical estimates based on closure, renormalization-group, and mode-coupling methods. The resulting semi-analytical parameters closely match the online-learned values obtained using ensemble Kalman inversion across several 2D geophysical turbulence setups~\cite[][]{guan2024online}. LES using these parameters reproduces key DNS statistics, including the tails of the vorticity distribution, and robustly outperforms dynamic Leith and Smagorinsky baselines.
\end{abstract}

\section{Introduction}
Subgrid-scale (SGS) closures are essential for LES of turbulent flows in the Earth system, where DNS is computationally infeasible \cite[][]{meneveau2000scale,pearson2017evaluation,ross2023benchmarking,bracco2025machine}. SGS closures are commonly categorized as structural or functional models \cite[]{sagaut2006large}. Structural closures can accurately represent SGS stress structures but may suffer from numerical instabilities and filter dependence \cite[]{leonard1975energy,clark1979evaluation,zanna2020data,jakhar2023learning}. Our recent work showed that including higher-order terms can yield stable and accurate LES of 2D geophysical turbulence \cite[]{jakhar2026analytical}. However, this approach requires high LES resolution. Functional closures, including eddy-viscosity and backscattering models, target the net energy and/or enstrophy transfer across the filter and can be effective at lower LES resolutions \cite[]{smagorinsky1963general,leith1996stochastic,jansen2014parameterizing,jansen2015energy,perezhogin2019stochastic,perezhogin20202d,grooms2023backscatter}. Related closure-theory studies have developed scale-dependent eddy-viscosity and backscatter parameterizations for 2D and geophysical turbulence \cite[]{kraichnan1976eddy,frederiksen1997eddy,kitsios2013scaling}.

Unlike the 3D Smagorinsky coefficient, which can be related to the Kolmogorov constant through a Lilly-type argument, the corresponding constants for 2D geophysical LES closures have remained largely empirical. Data-driven approaches have recently been used to optimize LES closure parameters from statistical data \cite[]{mons2021ensemble}. In our recent work, we used ensemble Kalman inversion (EKI) to learn optimal closure parameters for eight setups of 2D geophysical turbulence~\citep{guan2024online}. The optimized parameters were found to be nearly constant across cases with different flow regimes, and LES using these parameters outperformed common and dynamic-closure baselines. This raises a natural question: are the learned near-universal parameters merely empirical, or can they be explained by classical 2D turbulence theory?

In this note, we present, to the best of our knowledge, the first semi-analytical framework for deriving closure parameters for 2D geophysical turbulence from turbulence scaling. This framework also provides a physical interpretation of the near-universal parameters previously obtained via EKI-based online learning~\cite[][]{guan2024online}. Specifically, we semi-analytically estimate $C_\text{L}$, $C_\text{S}$, and $C_\text{JH}$ from the turbulent kinetic energy (TKE) direct-cascade scaling law
$\hat{E}(k)=A\eta^{2/3}k^{-3}$, where $k$ is wavenumber, $\eta$ is the enstrophy dissipation rate, and $A$ is an amplitude constant~\cite[]{kraichnan1967inertial,leith1968diffusion,batchelor1969computation}. We further derive the corresponding coefficients for the log-corrected spectrum
$\hat{E}(k)=A'\eta^{2/3}k^{-3}[\ln(k/k_f)]^{-1/3}$ and for a general power law
$\hat{E}(k)=A''\eta^{2/3}(k_*)^{p-3}k^{-p}$, with details given in the supplementary material. Since $A$, $A'$, and $A''$ are obtained from DNS or high-fidelity spectra, the approach is semi- rather than fully analytical.
As discussed later, estimates of the amplitude constants $A$, $A'$, and $A''$ from DNS agree with previous theoretical works~\citep{kraichnan1967inertial,kraichnan1971inertial,olla1991renormalization,nandy1995mode}. The resulting parameters closely match the EKI-optimized values and yield LES that reproduce key DNS statistics, suggesting that the previously learned optimal parameters are not merely empirical fits but can be interpreted through classical 2D turbulence theory.


\section{DNS, LES, and Closures}
We consider the dimensionless 2D $\beta$-plane vorticity equation \cite[e.g.,][]{vallis2017atmospheric}
\begin{eqnarray}\label{eq:NS}
\frac{\partial \omega}{\partial t}
+\mathcal{J}(\omega,\psi)
=
\frac{1}{\Rey}\nabla^2\omega
-f-r\omega
+\beta\frac{\partial \psi}{\partial x},
\qquad
\nabla^2\psi=-\omega .
\end{eqnarray}
Here, $\mathcal{J}$ is the Jacobian, $f(x,y)=k_f[\cos(k_fx)+\cos(k_fy)]$ is a time-independent deterministic forcing at wavenumber $k_f$, $\Rey$ is the Reynolds number, $\beta$ is the planetary-vorticity-gradient parameter, and $r=0.1$ is the linear friction coefficient.

Applying a sharp spectral cut-off filter, denoted by $\overline{(\cdot)}$, gives the LES or filtered-DNS (FDNS) equation
\begin{eqnarray}\label{eq:FNS}
\frac{\partial \overline{\omega}}{\partial t}
+\mathcal{J}(\overline{\omega},\overline{\psi})
=
\frac{1}{\Rey}\nabla^2\overline{\omega}
-\overline{f}
-r\overline{\omega}
+\beta\frac{\partial \overline{\psi}}{\partial x}
-\underbrace{
\left[
\overline{\mathcal{J}(\omega,\psi)}
-
\mathcal{J}(\overline{\omega},\overline{\psi})
\right]
}_{\Pi^\text{SGS}=\nabla\times(\nabla\cdot\tau^\text{SGS})},
\;
\nabla^2\overline{\psi}=-\overline{\omega}.
\end{eqnarray}
The SGS term $\Pi^\text{SGS}$ stems from the Jacobian and must be closed in terms of the resolved variables $(\overline{\psi},\overline{\omega})$. For eddy-viscosity closures,
\begin{eqnarray}
\tau^\text{SGS}=-2\nu_e\overline{\mathcal{S}},
\end{eqnarray}
where $\overline{\mathcal{S}}$ is the resolved strain-rate tensor. The Smagorinsky model uses
\begin{eqnarray}\label{eq:Smag}
\nu_e=(C_\text{S}\Delta)^2
\langle\overline{\mathcal{S}}^2\rangle^{1/2},
\end{eqnarray}
while the Leith model uses
\begin{eqnarray}\label{eq:Leith}
\nu_e=(C_\text{L}\Delta)^3
\langle(\nabla\overline{\omega})^2\rangle^{1/2}.
\end{eqnarray}
Here, $\Delta=L/N_\text{LES}$ is the filter width, and $\langle\cdot\rangle$ denotes domain averaging. The JH closure model is written as \cite[]{jansen2014parameterizing}
\begin{eqnarray}\label{eq:JH}
\Pi^\text{SGS}
=
\nabla^2(\nu_e\nabla^2\overline{\omega})
+
\nu_\text{B}\nabla^2\overline{\omega},
\end{eqnarray}
where the first term dissipates enstrophy through biharmonic eddy viscosity and the second term re-injects energy by anti-diffusion. Similar to their original papers~\citep{jansen2014parameterizing,jansen2015energy}, we use
\begin{eqnarray}\label{eq:nuB}
\nu_e
=
(C_\text{JH}\Delta)^6
\langle(\nabla^2\overline{\omega})^2\rangle^{1/2}, \;
\nu_\text{B}
=
-C_\text{B}
\frac{
\left\langle
\overline{\psi}\nabla^2(\nu_e\nabla^2\overline{\omega})
\right\rangle
}{
\left\langle
\overline{\psi}\nabla^2\overline{\omega}
\right\rangle
}.
\end{eqnarray}
The parameter $C_\text{B}$ prescribes the fraction of dissipated energy re-injected into the resolved scales. Following \citep{jansen2014parameterizing, guan2024online}, we set $C_\text{B}=0.95$ in all cases. Below, we derive the semi-analytical expression for scalar coefficients $C_\text{L}$, $C_\text{S}$, and $C_\text{JH}$.

\section{Semi-analytical Derivation of the Closure Parameters}
The derivation below is, to the best of our knowledge, the first Lilly-type global scaling argument for estimating scalar coefficients in commonly used 2D eddy-viscosity and backscattering closures. We treat the eddy viscosity as spatially uniform, $\nu_e=\nu_e(t)$, so the resulting coefficients should be interpreted as scalar, domain-averaged closure constants rather than local eddy-viscosity fields. The cutoff condition $\Delta\sim L_\eta$ is imposed up to a proportionality constant that is absorbed into the closure coefficient. A sharp spectral cutoff, $k_c=\pi/\Delta$, is used throughout. Detailed derivations and extensions to a general $k^{-p}$ scaling law are given in the supplementary material.

\subsection{$k^{-3}$ spectrum scaling}\label{sec:neg3_scale}

In 2D turbulence, the interscale enstrophy transfer is
\begin{eqnarray}
\eta=\langle\bar{\omega}\Pi\rangle,
\end{eqnarray}
where $\eta>0$ denotes transfer from resolved to subgrid or dissipative scales. For an eddy-viscosity closure $\Pi=-\nu_e\nabla^2\bar{\omega}$ with spatially uniform $\nu_e(t)$,
\begin{eqnarray}\label{eq:eta}
\eta=-\nu_e\langle\bar{\omega}\nabla^2\bar{\omega}\rangle
=\nu_e\langle(\nabla\bar{\omega})^2\rangle .
\end{eqnarray}
Requiring $\Delta$ to resolve the enstrophy-dissipation scale
$L_\eta\sim\nu_e^{1/2}\eta^{-1/6}\sim\nu_e^{1/3}
\langle(\nabla\bar{\omega})^2\rangle^{-1/6}$
\cite[]{batchelor1969computation,boffetta2007energy,boffetta2010evidence}
gives the Leith form
\begin{eqnarray}\label{eq:nuLeith}
\nu_e=(C_\text{L}\Delta)^3\langle(\nabla\bar{\omega})^2\rangle^{1/2}.
\end{eqnarray}
Combining Eqs.~\eqref{eq:eta} and \eqref{eq:nuLeith} yields
\begin{eqnarray}\label{eq:eta_leith}
\eta=(C_\text{L}\Delta)^3\langle(\nabla\bar{\omega})^2\rangle^{3/2}.
\end{eqnarray}
Using the direct-cascade TKE spectrum
\begin{eqnarray}\label{eq:TKE}
\hat{E}(k)=A\eta^{2/3}k^{-3}
\end{eqnarray}
\cite[]{kraichnan1967inertial,leith1968diffusion,batchelor1969computation}
and Parseval's theorem,
\begin{eqnarray}\label{eq:grad_vor_sq}
\langle(\nabla\bar{\omega})^2\rangle
=2\int_0^{k_c}k^4\hat{E}(k)\,dk
=A\eta^{2/3}k_c^2,
\end{eqnarray}
where $k_c=\pi/\Delta$. Here and below, the scaling law is applied to the FDNS spectrum up to $k_c$. For moments dominated by high wavenumbers, the error from extending the direct-cascade form below $k_f$ is assumed small. Substitution into Eq.~\eqref{eq:eta_leith} gives
\begin{eqnarray}\label{eq:CL}
C_\text{L}=\pi^{-1} A^{-1/2}.
\end{eqnarray}
Since \(C_\text{L}\) is determined up to the amplitude constant \(A\), the estimate is semi-analytical. To the best of our knowledge, Eq.~\eqref{eq:CL} gives the first explicit relation between a 2D LES closure coefficient and the amplitude constant of an enstrophy-cascade scaling law. Below, the same framework is extended to Smagorinsky, biharmonic eddy-viscosity, and JH backscattering closures. For the biharmonic eddy-viscosity part of the JH closure, setting $C_\text{B}=0$,
\begin{eqnarray}\label{eq:eta_hyper}
\eta=\nu_e\langle\bar{\omega}\nabla^4\bar{\omega}\rangle
=\nu_e\langle(\nabla^2\bar{\omega})^2\rangle,
\qquad
\nu_e=(C_\text{JH}\Delta)^6
\langle(\nabla^2\bar{\omega})^2\rangle^{1/2}.
\end{eqnarray}
The corresponding spectral moment is
\begin{eqnarray}\label{eq:grad_2_vor_sq}
\langle(\nabla^2\bar{\omega})^2\rangle
=2\int_0^{k_c}k^4\hat{Z}(k)\,dk
=\frac{A}{2}\eta^{2/3}k_c^4.
\end{eqnarray}
Substituting back into Eq.~\eqref{eq:eta_hyper} and using $k_c=\pi/\Delta$ gives
\begin{eqnarray}\label{eq:CHL}
C_\text{JH}=(A/2)^{-1/4}\pi^{-1}.
\end{eqnarray}

For the full JH model with a prescribed $C_\text{B}$, using Eq.~\eqref{eq:JH} in Eq.~\eqref{eq:eta} gives
\begin{eqnarray}\label{eq:eta_hyper_backscatter}
\eta=(C_\text{JH}\Delta)^6(A/2)^{3/2}\eta k_c^6
-\nu_\text{B}A\eta^{2/3}k_c^2.
\end{eqnarray}
From Eq.~\eqref{eq:nuB},
\begin{eqnarray}\label{eq:CB}
\nu_\text{B}
=C_\text{B}\nu_e
\frac{2\int_0^{k_c}k^2\hat{Z}(k)\,dk}
{2\int_0^{k_c}\hat{Z}(k)\,dk}
\approx
C_\text{B}\nu_e\frac{k_c^2}{2\ln(k_c)} ,
\end{eqnarray}
where the denominator is evaluated from the smallest nonzero wavenumber, $k=1$, in the $2\pi$-periodic domain. This logarithmic factor introduces a weak dependence on the large-scale cutoff and on the sharp-filter convention. Substituting Eq.~\eqref{eq:CB} into Eq.~\eqref{eq:eta_hyper_backscatter} gives
\begin{eqnarray}\label{eq:CHL_CB}
C_\text{JH}
=(A/2)^{-1/4}\pi^{-1}
\left(1-\frac{C_\text{B}}{\ln(k_c)}\right)^{-1/6}.
\end{eqnarray}

Finally, we estimate $C_\text{S}$ by matching the globally averaged Smagorinsky and Leith eddy viscosities for a resolved field with the assumed spectrum:
\begin{eqnarray}\label{eq:nu_e_being_equal}
(C_\text{L}\Delta)^3\langle(\nabla\bar{\omega})^2\rangle^{1/2}
=(C_\text{S}\Delta)^2\langle\bar{\mathcal{S}}^2\rangle^{1/2}.
\end{eqnarray}
With
\begin{eqnarray}\label{eq:S_sq}
\langle\bar{\mathcal{S}}^2\rangle
=2\int_0^{k_c}k^2\hat{E}(k)\,dk
\approx2A\eta^{2/3}\ln(k_c),
\end{eqnarray}
we obtain
\begin{eqnarray}\label{eq:CS}
C_\text{S}=A^{-3/4}\pi^{-1}(2\ln(k_c))^{-1/4}.
\end{eqnarray}
Equations~\eqref{eq:CL}, \eqref{eq:CHL_CB}, and \eqref{eq:CS} give semi-analytical expressions for the Leith, JH, and Smagorinsky coefficients under the classical \(k^{-3}\) enstrophy-cascade scaling, all determined up to the amplitude constant \(A\). To the best of our knowledge, this provides the first explicit scaling-based estimate of these commonly used 2D LES closure coefficients. We next repeat the derivation using the logarithmically corrected \(k^{-3}\) spectrum.
\subsection{$k^{-3}[\ln(k/k_f)]^{-1/3}$ spectrum scaling}\label{sec:log_scale}

For the logarithmically corrected direct-cascade spectrum \cite[]{kraichnan1971inertial}
\begin{eqnarray}\label{eq:TKE_log_correction}
\hat{E}(k)=A'\eta^{2/3}k^{-3}[\ln(k/k_f)]^{-1/3},
\end{eqnarray}
the scaling is applied only for $k>k_f$; in the discrete spectra we therefore use $k_f+1$ as the lower limit. The leading-order spectral moments are
\begin{eqnarray}
\langle(\nabla\bar{\omega})^2\rangle
&\approx&
2\int_{k_f+1}^{k_c}
A'\eta^{2/3}k[\ln(k/k_f)]^{-1/3}\,dk \nonumber\\
&\approx&
A'\eta^{2/3}k_c^2[\ln(k_c/k_f)]^{-1/3},
\label{eq:grad_vor_sq_log}
\end{eqnarray}
and
\begin{eqnarray}
\langle(\nabla^2\bar{\omega})^2\rangle
&\approx&
2\int_{k_f+1}^{k_c}
A'\eta^{2/3}k^3[\ln(k/k_f)]^{-1/3}\,dk \nonumber\\
&\approx&
\frac{A'}{2}\eta^{2/3}k_c^4[\ln(k_c/k_f)]^{-1/3}.
\label{eq:grad_2_vor_sq_log}
\end{eqnarray}
These approximations follow from the slow variation of the logarithmic factor, or equivalently from Karamata's theorem \cite[]{bingham1989regular}. Repeating the balances above gives
\begin{eqnarray}\label{eq:CL_log}
C'_\text{L}
=\frac{1}{\pi A'^{1/2}}[\ln(k_c/k_f)]^{1/6},
\end{eqnarray}
\begin{eqnarray}\label{eq:CJH_log}
C'_\text{JH}
=(A'/2)^{-1/4}\pi^{-1}[\ln(k_c/k_f)]^{1/12}.
\end{eqnarray}
Here, we use $C'_\text{L}$ and $C'_\text{JH}$ with $'$ to denote coefficients derived from the logarithmically corrected direct-cascade spectrum. For a prescribed $C_\text{B}$, the same asymptotic estimate gives
$\nu_\text{B}\approx (C_\text{B}/3)\nu_e k_c^2[\ln(k_c/k_f)]^{-1}$, leading to
\begin{eqnarray}\label{eq:CHL_CB_log}
C'_\text{JH}
=(A'/2)^{-1/4}\pi^{-1}[\ln(k_c/k_f)]^{1/12}
\left(1-\frac{2C_\text{B}}{3\ln(k_c/k_f)}\right)^{-1/6}.
\end{eqnarray}
Similarly, matching the globally averaged Smagorinsky and Leith eddy viscosities yields
\begin{eqnarray}\label{eq:CS_log}
C'_\text{S}=3^{-1/4}A'^{-3/4}\pi^{-1}.
\end{eqnarray}
\subsection{Effective \(k^{-4}\) scaling for the \(\beta\)-plane case}
\label{sec:p4_scale}

For the anisotropic \(\beta\)-plane case considered below, the angle-integrated TKE spectrum is steeper than the classical \(k^{-3}\) direct-cascade spectrum, as shown in Fig.~\ref{fig:1 DNS and FDNS}. We therefore also use an effective \(k^{-4}\) scaling,
\begin{eqnarray}
    \hat{E}(k)=A''\eta^{2/3}k_* k^{-4},
\end{eqnarray}
where \(k_*\) is a characteristic wavenumber used to keep \(A''\) dimensionless, and the choice of $k_*$ will affect the value of $A''$, but will not affect the derived closure coefficients. Following the same arguments and assumptions as above, and using the leading-order spectral moments
\begin{eqnarray}
\langle(\nabla\bar{\omega})^2\rangle
\approx 2A''\eta^{2/3}k_* k_c,
\qquad
\langle(\nabla^2\bar{\omega})^2\rangle
\approx \frac{2A''}{3}\eta^{2/3}k_* k_c^3,
\end{eqnarray}
we obtain the Leith coefficient and JH coefficient with prescribed \(C_\text{B}\)
\begin{eqnarray}
C''_\text{L}
=
\frac{1}{\pi}
\left(\frac{k_c}{2A''k_*}\right)^{1/2},\:
C''_\text{JH}
=
\frac{1}{\pi}
\left(\frac{2A''k_*}{3}\right)^{-1/4}
k_c^{1/4}
\left(1-\frac{3C_\text{B}}{k_c}\right)^{-1/6}.
\end{eqnarray}
Here, we use $C''_\text{L}$ and $C''_\text{JH}$ with $''$ to denote coefficients derived from the effective \(k^{-4}\) scaling. These expressions are used only as an effective isotropic representation of the angle-integrated spectrum in the anisotropic \(\beta\)-plane case. The general \(k^{-p}\) derivation is given in the supplementary material.

\section{Numerical Results} \label{sec:Results}
\begin{figure}
\vspace{.0in}
 \centering
 \vspace*{2mm}
 \begin{overpic}[width=1\linewidth,height=0.3\linewidth]{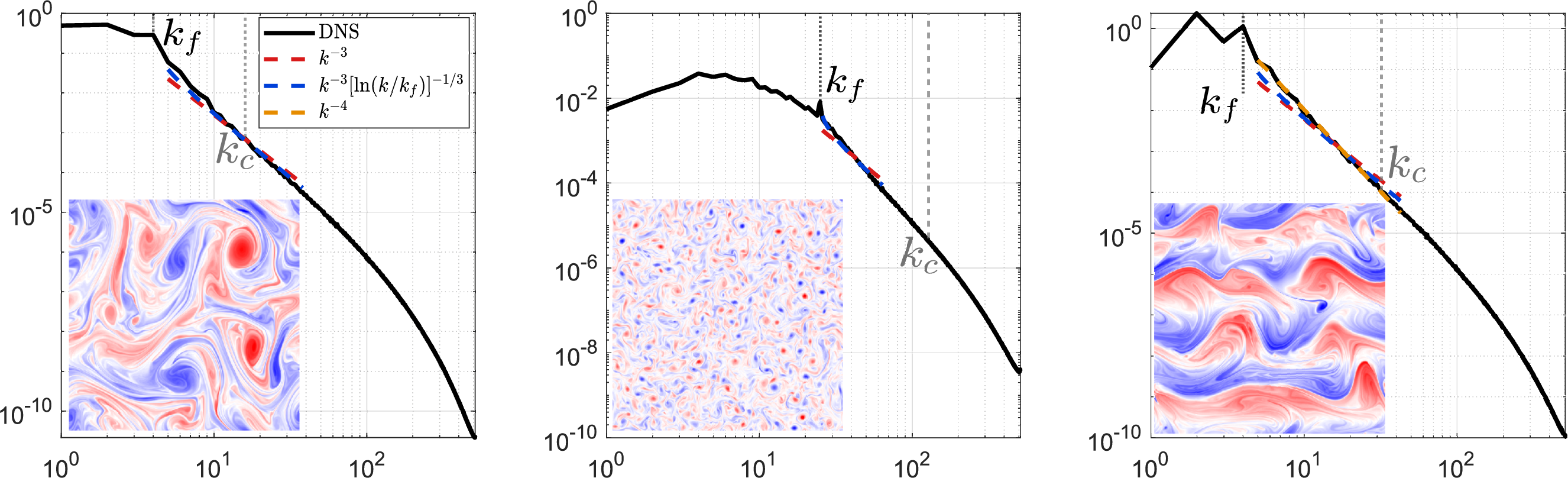}
 \put(11,30.5){{Case 1.1}}
 \put(45,30.5){{Case 2.1}}
 \put(81,30.5){{Case 3}}

 \put(-1,20){\scriptsize $\hat{E}(k)$}
 \put(33,20){\scriptsize $\hat{E}(k)$}
 \put(68,20){\scriptsize $\hat{E}(k)$}

\put(15,-1.0){\scriptsize $k$}
\put(50,-1.0){\scriptsize $k$}
\put(85,-1.0){\scriptsize $k$}
 
 \end{overpic}
  \vspace*{-4mm}
  \caption{\footnotesize The TKE spectra ($\hat{E}(k)$) and examples of $\omega$ for representative cases 1.1, 2.1, and 3. The $k_f$ and $k_c$ are marked. The black curves show DNS spectra averaged over 100 snapshots. The red (blue) dashed line shows the $k^{-3}$ scaling law  (with logarithmic correction), with the $A$ ($A'$) values shown in Table~\ref{tab:1}. The orange dashed line shows the effective $k^{-4}$ scaling law for Case 3 only. With a sharp spectral cut-off filter, the FDNS spectrum overlaps with the DNS spectrum over $k=[0,k_c]$. }
 \label{fig:1 DNS and FDNS}
 \vspace*{0mm}
\end{figure}

To evaluate the semi-analytical estimates of $C_\text{L}$, $C_\text{JH}$, and $C_\text{S}$, we use the DNS and LES datasets generated in~\cite{guan2024online}. The physical and numerical parameters of the eight cases are summarized in Table~\ref{tab:1}. The DNS TKE spectra and representative vorticity fields for three cases are shown in Fig.~\ref{fig:1 DNS and FDNS}.
Table~\ref{tab:1} also reports the diagnosed amplitudes $A$ and $A'$ obtained by fitting
$\hat{E}(k)=A\eta^{2/3}k^{-3}$ and
$\hat{E}(k)=A'\eta^{2/3}k^{-3}[\ln(k/k_f)]^{-1/3}$
to the direct-cascade range $k\in[k_f+1,k_\eta]$ of the DNS TKE spectra. In 2D turbulence, the enstrophy-cascade dissipation wavenumber scales as
$k_\text{diss}\sim \nu^{-1/2}\eta^{1/6}\sim Re^{1/2}\eta^{1/6}$~\cite[][]{boffetta2007energy,boffetta2010evidence}.
To avoid fitting the viscous-dissipation range, we set the fitting upper bound to
$k_\eta=\xi Re^{1/2}\eta^{1/6}$ with $\xi=0.25$.
Here, $\xi$ is an operational safety factor, chosen conservatively and tested by sensitivity analysis in the supplementary material. The enstrophy dissipation rate is diagnosed from the same DNS TKE spectra as~\cite[][]{davidson2015turbulence}
\begin{eqnarray}
    \eta=\langle(\nabla \omega)^2\rangle/Re=\frac{2}{Re}\int_0^{k_\text{DNS}} k^4\hat{E}(k)dk.
\end{eqnarray}
After the flow reaches statistical equilibrium, the snapshot-averaged spectrum and the diagnosed mean $\eta$ are statistically stationary. The TKE spectrum is calculated by averaging spectra from 100 decorrelated DNS snapshots. As discussed in \cite{guan2024online}, even one snapshot can be sufficient for estimating $A$ and $A'$ accurately. Except for Case~3, $A$ and $A'$ are nearly flow independent and are both close to $2$. The somewhat larger $A$ in Case~2.1 is likely related to the shorter available fitting interval between $k_f+1$ and $k_\eta$, making the original $k^{-3}$ scaling more sensitive to the selected fitting upper bound. This motivates the log-corrected scaling, which shows reduced sensitivity to $\xi$ in the supplementary material. The diagnosed values of $A'$ are close to the mode-coupling estimates of the log-corrected 2D enstrophy-cascade spectral constant reported by \cite{nandy1995mode}, including the corrected one-loop value $1.923$ and the self-consistent value $2.201$, and are broadly consistent with other theoretical estimates based on test-field closure approximation and renormalization-group methods~\cite[e.g.,][]{kraichnan1971inertial,olla1991renormalization}. The diagnosed $A$ values are consistent with empirical effective $k^{-3}$ amplitudes reported in 2D turbulence~\cite[e.g.,][]{boffetta2012two,gupta2019energy}. Case~3 has larger fitted amplitudes ($A=2.8$ and $A'=2.9$), likely due to the strong anisotropy and jet-like structures induced by the $\beta$ effect. As shown in Fig.~\ref{fig:1 DNS and FDNS}, the $k^{-3}$ and log-corrected $k^{-3}$ spectra are too shallow for this case. We therefore also consider an effective $k^{-4}$ scaling of the angle-integrated spectrum for Case~3. This scaling should be interpreted as an empirical isotropic representation of the steepened spectrum in the anisotropic $\beta$-plane case, rather than as a universal $\beta$-plane scaling law~\cite[e.g.,][]{rhines1975waves,sukoriansky2009transport,galperin2008zonostrophic,galperin2010geophysical}. The corresponding $C_\text{L}$ and $C_\text{JH}$ estimates obtained from this scaling agree reasonably well with the EKI-optimized values.

\begin{figure}
\vspace{.0in}
 \centering
 \vspace*{2mm}
 \begin{overpic}[width=0.8\linewidth,height=0.4\linewidth]{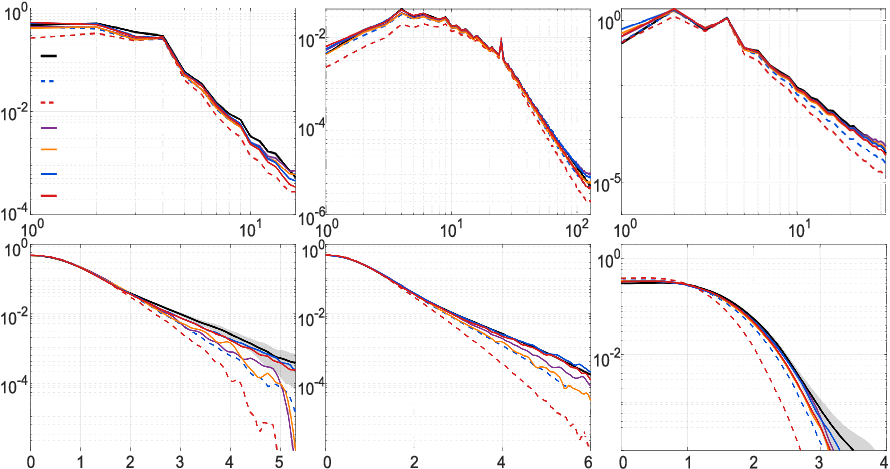}

 \put(11,50){{Case 1.1}}
 \put(45,50){{Case 2.1}}
 \put(81,50){{Case 3}}

\put(15,24.5){\footnotesize $k$}
 \put(54,24.5){\footnotesize $k$}
 \put(80,24.5){\footnotesize $k$}

  \put(15,-2){\footnotesize $\omega/\sigma_\omega$}
 \put(50,-2){\footnotesize $\omega/\sigma_\omega$}
 \put(80,-2){\footnotesize $\omega/\sigma_\omega$}

 \put(-8,10){\footnotesize $\mathcal{P}(\omega)$}
  \put(-8,35){\footnotesize $\hat{E}(k)$}

  \put(7.5,43.5){\scriptsize FDNS}
  \put(7.5,40.6){\scriptsize DLeith}
  \put(7.5,38.2){\scriptsize DSmag}
  \put(7.5,35.7){\scriptsize Leith ($C_\text{L}$)}
  \put(7.5,33.5){\scriptsize Leith ($C'_\text{L}$)}
  \put(7.5,30.7){\scriptsize JH ($C_\text{JH}$)}
  \put(7.5,28.4){\scriptsize JH ($C'_\text{JH}$)}
 
 \end{overpic}
  \vspace*{1mm}
  \caption{\footnotesize {\it A posteriori} testing with semi-analytical model parameters shown in Table~\ref{tab:1}. First row shows the TKE spectra ($\hat{E}(k)$) comparison among the ground truth (FDNS), baseline models (dynamic Leith, or DLeith, and dynamic Smagorinsky, or DSmag), and the semi-analytical models derived from the $k^{-3}$ scaling (with $C_\text{L}$ or $C_\text{JH}$) and the logarithmic correction (with $C'_\text{L}$ or $C'_\text{JH}$), for representative cases 1.1, 2.1, and 3. The semi-analytical models show a better match with the ground truth than DSmag among all 3 cases and better than DLeith in Case 3. The second row shows the positive tail of the vorticity probability density function, $\mathcal{P}(\omega)$. The negative tail gives the same qualitative conclusions because the vorticity is nearly symmetric about zero. The semi-analytical JH models consistently match the FDNS $\mathcal{P}(\omega)$ within 1 standard deviation (gray shaded area), especially in the tails or rare extreme events. The semi-analytical Leith models, although not as accurate as the JH models, outperform the baseline models.   
  }
 \label{fig:2 TKE and PDF}
 \vspace*{0mm}
\end{figure}

Overall, the semi-analytical estimates of $C_\text{L}$, $C_\text{JH}$, and $C_\text{S}$ agree well with the EKI-optimized values of \cite{guan2024online}. This agreement provides an independent validation of the semi-analytical approach and, furthermore, shows that the EKI-optimized parameters are not merely empirical fits but can be interpreted through classical 2D turbulence scaling. LES results using the semi-analytically estimated parameters for three representative cases are shown in Fig.~\ref{fig:2 TKE and PDF}. These cases are selected to represent distinct regimes and to include cases where the semi-analytical predictions deviate most from the EKI-optimized values. In a posteriori tests based on TKE spectra and vorticity probability density functions, LES with the semi-analytical parameters robustly outperform the dynamic Leith~\citep{maulik2016dynamic} and dynamic Smagorinsky~\citep{germano1991dynamic} baselines, particularly in the PDF tails. The backscattering JH closures also better capture the tails of the FDNS vorticity PDFs than purely diffusive Leith or Smagorinsky closures. More detailed a priori and a posteriori analyses of the flow statistics obtained with these closure parameters are reported in \cite{guan2024online}. The focus here is the semi-analytical derivation of the closure coefficients.
\begin{table*}
\centering
\small
\setlength{\tabcolsep}{3.0pt}
\renewcommand{\arraystretch}{1.18}
\newcommand{\val}[2]{\makecell[c]{#1\\(#2)}}
\caption{
Physical and numerical parameters of the eight cases and comparison between
semi-analytical predictions and EKI-optimized coefficients. The semi-analytical
predictions are obtained using either the $k^{-3}$ scaling or the logarithmically
corrected $k^{-3}[\ln(k/k_f)]^{-1/3}$ scaling. Values in parentheses denote one
standard deviation over the 100 DNS TKE spectra. For the JH closure,
$C_\text{JH}$ denotes the backscatter-corrected coefficient, computed with
$C_\text{B}=0.95$ for all semi-analytical predictions. For the anisotropic
$\beta$-plane Case 3, the effective $k^{-4}$ scaling predicts
$C_\text{L}=0.18 (0.0079)$ and $C_\text{JH}=0.32 (0.0070)$. Details of the DNS
setups and the $k^{-p}$ scaling-law prediction are provided in the supplementary
material.
}
\label{tab:1}
\begin{tabular}{lcccccccc}
\toprule
Case
& 1.1 & 1.2 & 1.3 & 1.4 & 2.1 & 2.2 & 2.3 & 3 \\
\midrule
\multicolumn{9}{l}{Numerical and physical setup} \\
$Re$
& 20000 & 20000 & 100000 & 300000 & 20000 & 100000 & 300000 & 20000 \\
$\beta$
& 0 & 0 & 0 & 0 & 0 & 0 & 0 & 20 \\
$k_f$
& 4 & 4 & 4 & 4 & 25 & 25 & 25 & 4 \\
$k_c$
& 16 & 32 & 128 & 128 & 128 & 128 & 128 & 32 \\

\midrule
\multicolumn{9}{l}{$k^{-3}$ prediction} \\
$A$
& \val{2.1}{0.21} & \val{2.1}{0.21} & \val{1.7}{0.017} & \val{1.8}{0.046}
& \val{2.7}{0.090} & \val{2.1}{0.013} & \val{1.9}{0.027} & \val{2.8}{0.28} \\
$C_\text{L}$
& \val{0.21}{0.010} & \val{0.21}{0.010} & \val{0.23}{0.0011} & \val{0.23}{0.0029}
& \val{0.19}{0.0031} & \val{0.21}{0.00066} & \val{0.22}{0.0015} & \val{0.18}{0.010} \\
$C_\text{S}$
& \val{0.11}{0.0087} & \val{0.11}{0.0082} & \val{0.11}{0.00085} & \val{0.11}{0.0021}
& \val{0.084}{0.0020} & \val{0.099}{0.00046} & \val{0.10}{0.0011} & \val{0.089}{0.0075} \\
$C_\text{JH}$
& \val{0.33}{0.0083} & \val{0.33}{0.0082} & \val{0.34}{0.00082} & \val{0.33}{0.0021}
& \val{0.30}{0.0025} & \val{0.32}{0.00049} & \val{0.33}{0.0011} & \val{0.30}{0.0083} \\

\midrule
\multicolumn{9}{l}{Log-corrected prediction} \\
$A'$
& \val{2.2}{0.21} & \val{2.2}{0.21} & \val{2.1}{0.020} & \val{2.4}{0.066}
& \val{2.0}{0.061} & \val{2.0}{0.010} & \val{2.0}{0.026} & \val{2.9}{0.27} \\
$C'_\text{L}$
& \val{0.22}{0.010} & \val{0.24}{0.011} & \val{0.26}{0.0013} & \val{0.25}{0.0035}
& \val{0.24}{0.0036} & \val{0.24}{0.00061} & \val{0.24}{0.0015} & \val{0.20}{0.010} \\
$C'_\text{S}$
& \val{0.13}{0.0095} & \val{0.13}{0.0095} & \val{0.13}{0.0010} & \val{0.12}{0.0026}
& \val{0.14}{0.0031} & \val{0.14}{0.00053} & \val{0.14}{0.0013} & \val{0.10}{0.0081} \\
$C'_\text{JH}$
& \val{0.35}{0.0084} & \val{0.35}{0.0084} & \val{0.35}{0.00087} & \val{0.34}{0.0024}
& \val{0.35}{0.0026} & \val{0.35}{0.00045} & \val{0.35}{0.0011} & \val{0.32}{0.0080} \\

\midrule
\multicolumn{9}{l}{EKI-optimized (online learned)} \\
$C^\text{EKI}_\text{L}$
& \val{0.23}{0.032} & \val{0.25}{0.028} & \val{0.26}{0.028} & \val{0.24}{0.025}
& \val{0.24}{0.026} & \val{0.23}{0.024} & \val{0.21}{0.035} & \val{0.21}{0.015} \\
$C^\text{EKI}_\text{S}$
& \val{0.12}{0.012} & \val{0.12}{0.010} & \val{0.11}{0.0041} & \val{0.12}{0.0082}
& \val{0.12}{0.008} & \val{0.12}{0.011} & \val{0.10}{0.015} & \val{0.10}{0.012} \\
$C^\text{EKI}_\text{JH}$
& \val{0.34}{0.019} & \val{0.32}{0.010} & \val{0.33}{0.0043} & \val{0.30}{0.0064}
& \val{0.32}{0.0051} & \val{0.32}{0.0036} & \val{0.31}{0.010} & \val{0.31}{0.014} \\
\bottomrule
\end{tabular}
\end{table*}

\section{Summary and Conclusion}\label{sec:conclusion}

We have developed semi-analytical estimates for the scalar coefficients in the Leith, Smagorinsky, and Jansen--Held closures for 2D geophysical turbulence. The derivation extends a Lilly-type global scaling argument to 2D turbulence and uses either the classical $k^{-3}$ direct-cascade spectrum or its log-corrected form above the forcing scale.

The estimates depend on an amplitude constant, e.g., $A$ or $A'$, and are therefore semi-analytical rather than fully analytical. For seven $\beta=0$ cases, however, the diagnosed amplitudes are nearly case-independent. In particular, the log-corrected amplitude is consistent with previous theoretical estimates based on test-field closure approximations, renormalization-group analysis, and mode-coupling theory. The resulting closure coefficients agree well with the EKI-optimized parameters obtained in our previous online-learning study \cite[][]{guan2024online}, suggesting that the learned optimal parameters are not merely empirical fits but can be interpreted through classical 2D turbulence scaling.

A posteriori LES tests show that closures using the semi-analytical parameters reproduce key DNS statistics and compare favorably with dynamic Leith and Smagorinsky baselines, especially in the vorticity-PDF tails. For the anisotropic $\beta$-plane case, the angle-integrated spectrum is steeper and is better represented by an effective $k^{-4}$ scaling, which should be viewed as a first approximation rather than a universal $\beta$-plane law. Future work will test these coefficients in more realistic geophysical models and develop an anisotropic theory for $\beta$-plane regimes. More broadly, together with the EKI results of \cite{guan2024online}, this work illustrates how modern data-driven closure discovery can motivate, guide, and validate new theoretical developments in turbulence modeling.

\vspace{5mm}
\noindent Declaration of Interests: The authors report no conflict of interest.
\vspace{-5mm}
\section*{Acknowledgments}
We thank Pavel Perezhogin for insightful discussions. This work was supported by NSF grant OAC-2005123, and by Schmidt Sciences, LLC. Computational resources were provided by ACCESS (ATM170020) and CISL (URIC0004).

\clearpage 

\setcounter{section}{0}
\setcounter{figure}{0}
\setcounter{table}{0}
\setcounter{equation}{0}

\renewcommand{\thesection}{S\arabic{section}}
\renewcommand{\thefigure}{S\arabic{figure}}
\renewcommand{\thetable}{S\arabic{table}}
\renewcommand{\theequation}{S\arabic{equation}}

\begin{center}
    \LARGE \textbf{Supplementary Materials}
\end{center}
\vspace{1cm}

In this online Supplementary Material,
\begin{enumerate}
    \item We present more details on the DNS setup and FDNS dataset.\\
    \item We provide the full detailed derivation of model parameters $C_\text{L}$, $C_\text{JH}$ without backscattering, $C_\text{JH}$ with backscattering, and $C_\text{S}$, using the $k^{-3}$ scaling law, the logarithmically corrected scaling law, and the general $k^{-p}$ scaling law.\\
    \item We perform a sensitivity analysis of the fitting range.
\end{enumerate}

\section{DNS setup and FDNS dataset}
To obtain target statistics, we solve Eqs.~(2.1) of the main text numerically. A Fourier-Fourier pseudo-spectral solver is used to solve these equations~\cite[e.g.,][]{guan2021stable,guan2023learning,subel2023explaining,jakhar2023learning,guan2024online,jakhar2026analytical}. The computational grid has uniform spacing $\Delta_\mathrm{DNS} = L/N_\mathrm{DNS}$, where $N_\mathrm{DNS}$ is the number of grid points in each direction. Here, cases 1.1, 1.2, 2.1, and 3 have $Re=20000$ and $N_\mathrm{DNS}=1024$; cases 1.3 and 2.2 have $Re=100000$ and $N_\mathrm{DNS}=4096$; and cases 1.4 and 2.3 have $Re=300000$ and $N_\mathrm{DNS}=4096$. Except for case 3 with $\beta=20$, other cases are isotropic with $\beta=0$. The time-stepping size is $\Delta t_\mathrm{DNS}= 5\times10^{-5}$ dimensionless time unit for cases 1.1, 1.2, 2.1, 3, and $\Delta t_\mathrm{DNS}= 1\times10^{-5}$ dimensionless time unit for cases 1.3, 1.4, 2.2, 2.3. For each case, using different random initial conditions, we conducted 5 independent DNS runs to generate the training statistics (from 4 runs) and {\it a posteriori} testing data sets (from 1 run). Once the flow reaches statistical equilibrium after a long-term spin-up, we uniformly sample $100$ snapshots that are $10000\Delta t_\mathrm{DNS}$ apart to obtain de-correlated data, following \cite{guan2023learning, guan2024online}.

To obtain the FDNS and to construct the training and testing data, we apply a sharp Fourier cut-off filter, with cut-off wavenumber $k_c=\pi/\Delta$ in each direction, generating $\overline{\psi}$, $\overline{\omega}$, and $\Pi$~\cite[e.g.,][]{pope2001turbulent,sagaut2006large}. This filtering and coarse-graining process is described in detail in~\cite{guan2021stable}. The DNS and FDNS spectra and flow fields of 3 representative cases (manifested with vorticity) are shown in the main text Fig.~1.

\section{Semi-analytical Derivation of the Closure Parameters}
The derivation below is a Lilly-type global scaling argument~\cite[][]{lilly1967representation} applied to 2D turbulence. We treat the eddy viscosity as spatially uniform, \(\nu_e=\nu_e(t)\), and therefore the resulting coefficients should be interpreted as scalar, domain-averaged closure constants. The cutoff condition \(\Delta\sim L_\eta\) is imposed up to a proportionality constant absorbed into the closure coefficient, and a sharp spectral cutoff \(k_c=\pi/\Delta\) is assumed. Here, we use the original $k^{-3}$ scaling law~\cite[]{kraichnan1967inertial,leith1968diffusion,batchelor1969computation} and its logarithmic correction~\cite[][]{kraichnan1971inertial}. We further extend the derivation to a general power law $\hat{E}(k)\sim k^{-p}$.

In 2D turbulence, the interscale (between resolved and subgrid scales) enstrophy transfer can be written as~\cite[]{thuburn2014cascades,guan2023learning}:
\begin{eqnarray}
\eta=\langle\bar{\omega}\Pi\rangle.
\end{eqnarray}
Here, we use the convention that $\eta>0$ is for enstrophy transfer from large scales to subgrid scales or dissipative scales.
\subsection{Leith eddy-viscosity model}\label{sec:leith}
When the SGS term is modeled by an eddy-viscosity model ($\Pi = -\nu_e\nabla^2\bar{\omega}$) with a spatially uniform but time-dependent $\nu_e(t)$~\cite[]{davidson2015turbulence},
\begin{eqnarray} \label{eq:eta}
\eta=-\nu_e\langle\bar{\omega}\nabla^2\bar{\omega}\rangle=\nu_e\langle(\nabla\bar{\omega})^2\rangle,
\end{eqnarray}
at a high-$Re$ turbulence flow where the molecular viscosity is much smaller than the eddy viscosity. The enstrophy-dissipation length scale for the eddy viscosity model is~\cite[]{batchelor1969computation,boffetta2007energy,boffetta2010evidence}
\begin{eqnarray}
    L_\eta(\nu_e) \sim \nu_e^{1/2} \eta^{-1/6} \sim \nu_e^{1/3}\langle(\nabla \bar{\omega})^2\rangle^{-1/6}.
\end{eqnarray}
In LES, the grid spacing, which is equal to the filter width $\Delta$ for sharp cut-off filtering, needs to resolve the enstrophy-dissipation length scale. Therefore:
\begin{eqnarray}
    \Delta = L_\eta(\nu_e) \sim \nu_e^{1/3}\langle(\nabla \bar{\omega})^2\rangle^{-1/6}.
\end{eqnarray}
Rearranging gives:
\begin{eqnarray} 
    \nu_e \sim \Delta^3\langle(\nabla \bar{\omega})^2\rangle^{1/2} =(C_\text{L}\Delta)^3\langle(\nabla \bar{\omega})^2\rangle^{1/2},
    \label{eq:nuLeith}
\end{eqnarray}
which is the Leith model with constant of proportionality $C_\text{L}$, although Leith initially derived this model from the Smagorinsky model using dimensional analysis~\cite[]{leith1996stochastic}. Substituting Eq.~\eqref{eq:nuLeith} into Eq.~\eqref{eq:eta} yields:
\begin{eqnarray} \label{eq:eta_leith}
\eta=(C_\text{L}\Delta)^3\langle(\nabla \bar{\omega})^2\rangle^{3/2}.
\end{eqnarray}
The term $\langle(\nabla \bar{\omega})^2\rangle$ can be written in terms of the enstrophy spectra, according to Parseval's theorem and the direct-cascade scaling law:
\begin{eqnarray}
 \hat{E}(k)=A\eta^{2/3}k^{-3}, \label{eq:TKE}   
\end{eqnarray}
where $A$ is a constant that can depend on the flow~\cite[]{kraichnan1967inertial,leith1968diffusion,batchelor1969computation}:
\begin{eqnarray}
\langle(\nabla \bar{\omega})^2\rangle &=&\langle\left(\partial \bar{\omega}/\partial x\right)^2+\left(\partial \bar{\omega}/\partial y\right)^2\rangle =\langle (ik_x\hat{\bar{\omega}})^*(ik_x\hat{\bar{\omega}})+(ik_y\hat{\bar{\omega}})^*(ik_y\hat{\bar{\omega}})\rangle, \label{eq:grad_vor_sq1} \\
&=&\langle k^2\hat{\bar{\omega}}^*\hat{\bar{\omega}}\rangle=2\langle k^2\hat{Z}(k)\rangle=2\langle k^4\hat{E}(k)\rangle,\\
&=& 2\int_0^{k_c} k^4\hat{E}(k)dk =  2\int_0^{k_c} A\eta^{2/3}kdk =  A\eta^{2/3}k_c^2,\label{eq:grad_vor_sq}
\end{eqnarray}
where $k_c=\pi/\Delta$ is the LES sharp cut-off wavenumber. Here and throughout the rest of the supplementary material, we apply the scaling law only to up to $k_c$, i.e., to the filtered DNS (FDNS) spectrum. We also assume that the error in approximating the integral for $k<k_f$ with the $k^{-3}$ scaling law (rather than the $k^{-5/3}$ scaling law) is relatively small. Therefore, Eq.~\eqref{eq:eta_leith} becomes:
\begin{eqnarray}
\eta=(C_\text{L}\Delta)^3(A\eta^{2/3}k_c^2)^{3/2}.
\end{eqnarray}
Factoring out $\eta$ and using $k_c=\pi/\Delta$ gives:
\begin{eqnarray}\label{eq:CL}
C_\text{L}=1/(\pi A^{1/2}).
\end{eqnarray}
This equation only requires knowing $A$ to determine $C_\text{L}$.  

\subsection{Biharmonic eddy-viscosity model}\label{sec:JH}
Focusing only on the eddy-viscosity part of the JH model (by setting $C_\text{B}=0$) and assuming $\nu=\nu(t)$ (spatial uniformity), the enstrophy interscale transfer becomes:
\begin{eqnarray} \label{eq:eta_hyper}
\eta=\nu_e\langle\bar{\omega}\nabla^4\bar{\omega}\rangle=\nu_e\langle(\nabla^2\bar{\omega})^2\rangle.
\end{eqnarray}
The balance between the enstrophy interscale transfer and the dissipation was also discussed in~\cite[]{jansen2014parameterizing}. The enstrophy-dissipation length scale for biharmonic eddy-viscosity can be obtained by dimensional analysis:
\begin{eqnarray}
        \Delta = L_\eta(\nu_e) \sim  \nu_e^{3/12} \eta^{-1/12} \sim  \nu_e^{1/6}\langle(\nabla^2 \bar{\omega})^2\rangle^{-1/12}.
\end{eqnarray}
Rearranging gives:
\begin{eqnarray}
    \nu_e \sim \Delta^6\langle(\nabla^2 \bar{\omega})^2\rangle^{1/2} =(C_\text{JH}\Delta)^6\langle(\nabla^2 \bar{\omega})^2\rangle^{1/2}.
\end{eqnarray}
Similar to the analysis in Eqs.~\eqref{eq:grad_vor_sq1}-\eqref{eq:grad_vor_sq}, we obtain:
\begin{eqnarray}
\langle(\nabla^2\bar{\omega})^2\rangle &=& \langle(\nabla^2\bar{\omega})(\nabla^2\bar{\omega})\rangle = \langle (-k^2\hat{\bar{\omega}})^*(-k^2\hat{\bar{\omega}})\rangle = \langle k^4\hat{\bar{\omega}}^*\hat{\bar{\omega}}\rangle,\\
 &=& 2\int_0^{k_c} k^4 \hat{Z}(k)dk =  2A\eta^{2/3}\int_0^{k_c}k^{3}dk = \frac{A}{2}\eta^{2/3}k_c^{4}.\label{eq:grad_2_vor_sq}
\end{eqnarray}
Eq.~\eqref{eq:grad_2_vor_sq} is equivalent to Eq. (A.3) in~\cite[]{jansen2014parameterizing}. Therefore, Eq.~\eqref{eq:eta_hyper} becomes:
\begin{eqnarray}
\eta = \nu_e\langle(\nabla^2\bar{\omega})^2\rangle = (C_\text{JH}\Delta)^6\langle(\nabla^2 \bar{\omega})^2\rangle^{3/2} = (C_\text{JH}\Delta)^6\left(\frac{A}{2}\eta^{2/3}k_c^{4}\right)^{3/2}.\label{eq:etaJH}
\end{eqnarray}
Factoring out $\eta$ and using $k_c=\pi/\Delta$ gives:
\begin{eqnarray}\label{eq:CHL}
C_\text{JH} = (A/2)^{-1/4}\pi^{-1}.
\end{eqnarray}
This equation, like Eq.~\eqref{eq:CL}, only requires constant $A$ to determine $C_\text{JH}$. 
\subsection{Jansen-Held (JH) backscattering model}\label{sec:JHB}
Using Eq.~2.6 in the main text in \eqref{eq:eta} and assuming $\nu=\nu(t)$ yields:
\begin{eqnarray} \label{eq:eta_hyper_backscatter}
\eta = (C_\text{JH}\Delta)^6(A/2)^{3/2}\eta k_c^6 - \nu_\text{B}A\eta^{2/3} k_c^2,
\end{eqnarray}
where Eqs.~\eqref{eq:etaJH} and \eqref{eq:grad_vor_sq} are used in the first and second terms, respectively. Starting from the definition of $\nu_\text{B}$ (Eq.~2.8 in the main text), we obtain:    
\begin{eqnarray}
    \nu_\text{B} &=& -C_{\text{B}}\frac{\left\langle \bar{\psi}\nu_e \nabla^4\bar{\omega}\right\rangle}{\langle\bar{\psi}\nabla^2\bar{\omega}\rangle} 
= C_\text{B}\nu_e\frac{2\int_0^{k_c} k^2\hat{Z}(k)dk}{2\int_0^{k_c} \hat{Z}(k)dk} 
 \approx C_\text{B}\nu_e\frac{2\int_0^{k_c} k^2\hat{Z}(k)dk}{2\int_1^{k_c} \hat{Z}(k)dk}\label{eq:approximation_integral},\\
&\approx&C_\text{B}\nu_e\frac{A\eta^{2/3}k_c^2}{2\int_1^{k_c} A\eta^{2/3}k^{-1}dk}
 \approx C_\text{B}\nu_e\frac{k_c^2}{2(ln(k_c))}.\label{eq:CB}
\end{eqnarray}
The approximation in Eq.~\eqref{eq:approximation_integral} is made since in numerical integration in a doubly periodic domain of length $2\pi$, wavenumbers are integers and $\hat{Z}(k=0)=0$. Using this expression for $\nu_\text{B}$ in Eq.~\eqref{eq:eta_hyper_backscatter} gives:
\begin{eqnarray}
    \eta^\text{} &=& (C_\text{JH}\Delta)^6(A/2)^{3/2}\eta k_c^6 - AC_\text{B}(C_\text{JH}\Delta)^6(A/2)^{1/2}\eta k_c^6/(2ln(k_c)).\label{eq:eta_JH}
\end{eqnarray}
Factoring out $\eta$ gives:
\begin{eqnarray}\label{eq:CHL_CB}
    C_\text{JH}&=&(A/2)^{-1/4}\pi^{-1} (1-C_\text{B}/ln(k_c))^{-1/6}.
\end{eqnarray}
With backscattering ($C_\text{B}>0$), $C_\text{JH}$ increases. In the JH model, $C_\text{JH}$ also weakly depends on the LES resolution due to a logarithmic correction of $k_c$. $C_\text{JH}$ decreases with an increase of LES resolution.

\subsection{Relation between $C_\text{S}$ and $C_\text{L}$ for 2D turbulence}\label{sec:Cs_CL_relation}
As mentioned earlier, an analytical derivation for $C_\text{S}$ exists for 3D but not for 2D turbulence. Here, we derive the relation between $C_\text{S}$ and $C_\text{L}$, by analyzing the relationship between $\langle\bar{\mathcal{S}}^2\rangle^{1/2}$ and $\Delta{\langle(\nabla \bar{\omega})^2\rangle^{1/2}}$, which appear in Eqs.~2.4 and~2.5 in the main text. This will enable us to derive a semi-analytical equation for $C_\text{S}$ for 2D. 

Similar to the derivation for Eq.~\eqref{eq:grad_vor_sq}, we have:
\begin{eqnarray}
    \langle\bar{\mathcal{S}}^2\rangle &=& 2\int_0^{k_c} k^2 \hat{E}(k)dk
    \approx 2\int_1^{k_c} k^2\hat{E}(k)dk,\label{eq:approximation_Cs}\\
    &\approx & 2\int_1^{k_c} k^2A\eta^{2/3}k^{-3}dk \approx2A\eta^{2/3}ln(k_c).\label{eq:S_sq}
\end{eqnarray}
As in the approximation used for Eq.~\eqref{eq:CB}, here we integrate from wavenumber $k=1$. Combining Eqs.~\eqref{eq:grad_vor_sq} and~\eqref{eq:S_sq} gives the ratio:
\begin{eqnarray}
    \frac{\Delta \langle(\nabla\bar{\omega})^2\rangle^{1/2}}{\langle\bar{\mathcal{S}}^2\rangle^{1/2}} &\approx& \Delta\left(\frac{A\eta^{2/3}k_c^2}{2A\eta^{2/3}ln(k_c)}\right)^{1/2}=\frac{\Delta\sqrt{(\pi/\Delta)^2}}{\sqrt{2ln(k_c)}}=\frac{\pi}{\sqrt{2ln(k_c)}}.\label{eq:ratio_leith_smag}
\end{eqnarray}
Assuming that the globally averaged eddy viscosities from Smag and Leith are matched for a resolved field with the assumed spectrum, we obtain:
\begin{eqnarray}
    \nu_e=(C_\text{L}\Delta)^3 \langle(\nabla\bar{\omega})^2\rangle^{1/2}=(C_\text{S}\Delta)^2\langle\bar{\mathcal{S}}^2\rangle^{1/2},\label{eq:nu_e_being_equal}
\end{eqnarray} 
Using Eqs.~\eqref{eq:ratio_leith_smag} and~\eqref{eq:nu_e_being_equal} gives:
\begin{eqnarray}
C_\text{S}=A^{-3/4} \pi^{-1} (2ln(k_c))^{-1/4}.\label{eq:CS}
\end{eqnarray}
Similar to $C_\text{JH}$, in 2D turbulence $C_\text{S}$ weakly depends on the LES resolution due to a logarithmic correction of $k_c$. $C_\text{S}$ decreases as $k_c$ (LES resolution) increases, given that $k_c$ is within the inertial range.

\subsection{$k^{-3}[ln(k/k_f)]^{-1/3}$ spectrum scaling}\label{sec:log_scale}
To better capture the inertial range, Kraichnan suggested a log correction to the original $k^{-3}$ law~\cite{kraichnan1971inertial}:
\begin{eqnarray}
 \hat{E}(k)=A'\eta^{2/3}k^{-3}[ln(k/k_f)]^{-1/3}, \label{eq:TKE_log_correction}   
\end{eqnarray}
where $A'$ is a constant and $k_f$ is the enstrophy injection wavenumber or forcing wavenumber in this case. Eq.~\eqref{eq:grad_vor_sq} becomes:
\begin{eqnarray}
\langle(\nabla \bar{\omega})^2\rangle 
= 2\int_0^{k_c} k^4\hat{E}(k)dk =  2\int_{k_f+1}^{k_c} A'\eta^{2/3}k[ln(k/k_f)]^{-1/3}dk.\label{eq:grad_vor_sq_log}
\end{eqnarray}
Since $[ln(k/k_f)]^{-1/3}$ is a slowly varying function of $k$, the integral is asymptotically dominated by its upper limit. Changing the integration variable $k$ to $k'=k/k_c$ and expanding in the power of $[ln(k_c/k_f)]^{-1}$ gives:
\begin{eqnarray}
\langle(\nabla \bar{\omega})^2\rangle &=& 2\int_{k_f+1}^{k_c} A'\eta^{2/3}k[ln(k/k_f)]^{-1/3}dk,\\
&=&A'\eta^{2/3}k_c^2[ln(k_c/k_f)]^{-1/3}[1+\mathcal{O}([ln(k_c/k_f)]^{-1})],\\
&\approx& A'\eta^{2/3}k_c^2[ln(k_c/k_f)]^{-1/3}.
\label{eq:grad_vor_sq_log2}
\end{eqnarray}
The approximation can also be obtained from the Karamata integration theorem~\cite[]{bingham1989regular}. Following the derivation shown in Sec.~\ref{sec:leith}, we have:
\begin{eqnarray}\label{eq:CL_log}
C'_\text{L}=1/(\pi A'^{1/2})[ln(k_c/k_f)]^{1/6}.
\end{eqnarray}
Similarly, Eq.~\eqref{eq:grad_2_vor_sq} becomes:
\begin{eqnarray}
\langle(\nabla^2\bar{\omega})^2\rangle &=& 2\int_0^{k_c} k^4 \hat{Z}=2\int_{k_f+1}^{k_c} A'\eta^{2/3}k^3[ln(k/k_f)]^{-1/3}dk,\\
&\approx& \frac{A'}{2}\eta^{2/3}k_c^4[ln(k_c/k_f)]^{-1/3}.
\label{eq:grad_2_vor_sq_log}
\end{eqnarray}
The approximation can also be obtained from the Karamata integration theorem for slow varying function $[ln(k/k_f)]^{-1/3}$. Following the discussion in Sec.~\ref{sec:JH}, we have:
\begin{eqnarray}\label{eq:CJH_log}
C'_\text{JH}=(A'/2)^{-1/4}\pi^{-1}[ln(k_c/k_f)]^{1/12}.
\end{eqnarray}
In case that backscattering is considered, Eq.~\eqref{eq:CB} becomes
\begin{eqnarray}\label{eq:CB_log}
    \nu_\text{B} &=& C_\text{B}\nu_e\frac{2\int_0^{k_c} k^2\hat{Z}(k)dk}{2\int_0^{k_c} \hat{Z}(k)dk} \approx C_\text{B}\nu_e\frac{\int_{k_f+1}^{k_c} A'\eta^{2/3}k[ln(k/k_f)]^{-1/3}dk}{\int_{k_f+1}^{k_c} A'\eta^{2/3}k^{-1}[ln(k/k_f)]^{-1/3}dk}, \\
    &=&C_\text{B}\nu_e\frac{\int_{k_f+1}^{k_c} k[ln(k/k_f)]^{-1/3}dk}{\int_{k_f+1}^{k_c} k^{-1}[ln(k/k_f)]^{-1/3}dk}
 \approx C_\text{B}\nu_e\frac{\frac{1}{2}k_c^2[ln(k_c/k_f)]^{-1/3}}{\frac{3}{2}[ln(k_c/k_f)]^{2/3}},\\ 
 &=&\frac{C_\text{B}}{3}\nu_ek_c^2[ln(k_c/k_f)]^{-1}.
\end{eqnarray}
For the numerator of Eq.~\eqref{eq:CB_log}, we use the Karamata integration theorem, and for the denominator, we use the same assumption as in Eq.~\eqref{eq:CB}. Similar to Eq.~\eqref{eq:eta_JH}-\eqref{eq:CHL_CB}, we have:
\begin{eqnarray}
    C'_\text{JH}&=&(A'/2)^{-1/4}\pi^{-1}[ln(k_c/k_f)]^{1/12} \bigg(1-\frac{2C_\text{B}}{3ln(k_c/k_f)}\bigg)^{-1/6}.\label{eq:CHL_CB_log}
\end{eqnarray}
With backscattering ($C_\text{B}>0$), $C_\text{JH}$ increases. In the JH model, $C_\text{JH}$ also weakly depends on the LES resolution due to a logarithmic correction of $k_c$. $C_\text{JH}$ decreases with an increase of LES resolution.  \\

The relationship between $C_\text{S}$ and $C_\text{L}$ can also be modified by:
\begin{eqnarray}
    \frac{\Delta \langle(\nabla\bar{\omega})^2\rangle^{1/2}}{\langle\bar{\mathcal{S}}^2\rangle^{1/2}} &\approx& \Delta\left(\frac{A'\eta^{2/3}k_c^2[ln(k_c/k_f)]^{-1/3}}{3A'\eta^{2/3}[ln(k_c/k_f)]^{2/3}}\right)^{1/2}=\frac{\Delta k_c}{[3ln(k_c/k_f)]^{1/2}},\\
    &=&\frac{\pi}{\sqrt{3ln(k_c/k_f)}}.\label{eq:ratio_leith_smag_log}
\end{eqnarray}
Following Eq.~\eqref{eq:nu_e_being_equal}, we have:
\begin{eqnarray}
    (C'_\text{S})^2 = (C'_\text{L})^3\frac{\pi}{\sqrt{3ln(k_c/k_f)}}.\label{eq:nu_e_being_equal_log}
\end{eqnarray} 
With Eq.~\eqref{eq:CL_log}, we have:
\begin{eqnarray}
    C'_\text{S}=3^{-1/4}A'^{-3/4}\pi^{-1}.\label{eq:CS_log}
\end{eqnarray} 

In summary, for a log-corrected $k^{-3}$ scaling law $\hat{E}(k)\sim k^{-3}[ln(k/k_f)]^{-1/3}$, $C'_\text{L}$, $C'_\text{S}$, and $C'_\text{JH}$ (with $C_\text{B}$) are functions of $A'$ and $k_c/k_f$. Here, $A'$ is the scaling coefficient that can be obtained from the DNS spectrum, and $k_f$ is the forcing wavenumber.\\

{\subsection{$k^{-p}$ spectrum scaling}}
For those cases whose TKE direct-cascade scaling law follows $k^{-p}$, e.g., $\hat{E}(k)=A''\eta^{2/3}(k_*)^{p-3}k^{-p}$ or equivalently $\hat{E}(k)=A''\eta^{2/3}k^{-3}(k/k_*)^{3-p}$, where $k_*$ is a characteristic wavenumber, we derive the relationships between $C''_\text{L}$, $C''_\text{S}$, $C''_\text{JH}$, and $C_\text{B}$ and $A''$ as follows. For a general case with moment $n$ and scaling exponent $p$, we have the TKE's n-momentum:
\begin{eqnarray}
    \int_0^{k_c} k^n\hat{E}(k)dk = A''\eta^{2/3}k_*^{p-3}\int_0^{k_c}k^{n-p}dk.\label{eq:TKE_n_momentum}
\end{eqnarray}
For $p< n+1$, the integrand $k^{n-p}$ is integrable at $k=0$, and the lower limit at $k=0$ is inconsequential. We can rewrite Eq.~\eqref{eq:TKE_n_momentum} by:
\begin{eqnarray}
    \int_0^{k_c} k^n\hat{E}(k)dk = A''\eta^{2/3}k_*^{p-3}\frac{k_c^{n-p+1}}{n-p+1}.\label{eq:TKE_n_momentum_approx_2}
\end{eqnarray}
For $p>n+1$, the integrand diverges at $k=0$. In a doubly periodic domain, the smallest nonzero wavenumber is $k_1=1$, and $\hat{E}(k=0)=0$. Therefore, we can approximate the integral by integrating from $k_1=1$:
\begin{eqnarray}
    \int_0^{k_c} k^n\hat{E}(k)dk \approx A''\eta^{2/3}k_*^{p-3}\int_{k_1}^{k_c}k^{n-p}dk=A''\eta^{2/3}k_*^{p-3}\frac{k_c^{n-p+1}-k_1^{n-p+1}}{n-p+1}.\label{eq:TKE_n_momentum_approx}
\end{eqnarray}
In the special case where $p=n+1$ (e.g., Eq.~\eqref{eq:S_sq} with $n=2, p=3$), we have:
\begin{eqnarray}
    \int_0^{k_c} k^n\hat{E}(k)dk \approx A''\eta^{2/3}k_*^{p-3}\int_{k_1}^{k_c}k^{n-p}dk=A''\eta^{2/3}k_*^{p-3}ln(k_c).\label{eq:TKE_n_momentum_approx_1}
\end{eqnarray}
In another special case when $p \geq n+2$, $k_c^{n-p+1} \leq 1/k_c << 1 = k_1^{n-p+1}$, we can approximate Eq.~\eqref{eq:TKE_n_momentum_approx} by:
\begin{eqnarray}
    \int_0^{k_c} k^n\hat{E}(k)dk \approx A''\eta^{2/3}k_*^{p-3}\frac{-k_1^{n-p+1}}{n-p+1}.\label{eq:TKE_n_momentum_approx_3}
\end{eqnarray}
Here, we keep $k_1$ instead of writing it out as $k_1=1$ to conserve dimensions and units. Using the equations Eqs.~\eqref{eq:TKE_n_momentum_approx_2}-~\eqref{eq:TKE_n_momentum_approx_3}, we can derive the relationships between $C''_\text{L}$, $C''_\text{S}$, $C''_\text{JH}$, and $C_\text{B}$ and $A''$. Here, we show two examples: $p<3$ and $p=4$.\\

\subsubsection{Relationships between $C''_\text{L}$, $C''_\text{S}$, $C''_\text{JH}$, and $C_\text{B}$ and $A''$ for $p < 3$}
Following the derivation in Secs.~\ref{sec:leith}-~\ref{sec:Cs_CL_relation}, we have:
\begin{eqnarray}
    \langle\bar{\mathcal{S}}^2\rangle &=& 2\int_0^{k_c} k^2 \hat{E}(k)dk
    =\frac{2A''}{3-p}\eta^{2/3}k_*^{p-3}k_c^{3-p}, \label{eq:S_sq_p}\\
     \langle(\nabla \bar{\omega})^2\rangle 
&=& 2\int_0^{k_c} k^4\hat{E}(k)dk =  \frac{2A''}{5-p}\eta^{2/3}k_*^{p-3}k_c^{5-p},\label{eq:grad_vor_sq_p}\\
\langle(\nabla^2 \bar{\omega})^2\rangle 
&=& 2\int_0^{k_c} k^6\hat{E}(k)dk = \frac{2A''}{7-p}\eta^{2/3}k_*^{p-3}k_c^{7-p}.\label{eq:grad_2_vor_sq_p}
\end{eqnarray}
Note that, if $p=3$, $\langle\bar{\mathcal{S}}^2\rangle$ will take the form as in Eq.~\eqref{eq:S_sq}, Eq.~\eqref{eq:grad_vor_sq_p} will be the same as Eq.~\eqref{eq:grad_vor_sq}, and Eq.~\eqref{eq:grad_2_vor_sq_p} will be the same as Eq.~\eqref{eq:grad_2_vor_sq}.\\

With Eqs.~\eqref{eq:S_sq_p}-~\eqref{eq:grad_2_vor_sq_p}, we can derive $C''_\text{L}$, $C''_\text{S}$, and $C''_\text{JH}$ (with $C_\text{B}$), accordingly. For example, Eq.~\eqref{eq:eta_leith} gives:
\begin{eqnarray}
    \eta=(C''_\text{L}\Delta)^3\bigg(\frac{2A''}{5-p}\eta^{2/3}k_*^{p-3}k_c^{5-p}\bigg)^{3/2}=(C''_\text{L}\Delta)^3\bigg( \frac{2A''}{5-p}\bigg)^{3/2}\eta k_*^{\frac{3}{2}(p-3)}k_c^{\frac{3}{2}(5-p)}.
\end{eqnarray}
Factoring out $\eta$ and using $k_c=\pi/\Delta$ gives:
\begin{eqnarray}
    C''_\text{L} = \frac{1}{\pi}\bigg(\frac{5-p}{2A''}\bigg)^{1/2}\bigg(\frac{k_c}{k_*}\bigg)^{\frac{1}{2}(p-3)}.\label{eq:CL_p}
\end{eqnarray}
Similarly, Eq.~\eqref{eq:eta_hyper_backscatter} becomes:
\begin{eqnarray}
\eta = \nu_e\frac{2A''}{7-p}\eta^{2/3}k_*^{p-3}k_c^{7-p} - \nu_\text{B}\frac{2A''}{5-p}\eta^{2/3}k_*^{p-3}k_c^{5-p}. \label{eq:etaJH_p}
\end{eqnarray}
Further, similar to Eq.~\eqref{eq:CB}, we have:
\begin{eqnarray}
    \nu_\text{B} &=& -C''_{\text{B}}\frac{\left\langle \bar{\psi}\nu_e \nabla^4\bar{\omega}\right\rangle}{\langle\bar{\psi}\nabla^2\bar{\omega}\rangle} 
= C_\text{B}\nu_e\frac{2\int_0^{k_c} k^2\hat{Z}(k)dk}{2\int_0^{k_c} \hat{Z}(k)dk}= C_\text{B}\nu_e\frac{2\int_0^{k_c} k^4\hat{E}(k)dk}{2\int_0^{k_c}k^2 \hat{E}(k)dk},\\
&=&C_\text{B}\nu_e\frac{\frac{2A''}{5-p}\eta^{2/3}k_*^{p-3}k_c^{5-p}}{\frac{2A''}{3-p}\eta^{2/3}k_*^{p-3}k_c^{3-p}} = C_\text{B}\nu_e\frac{3-p}{5-p}k_c^2.
\label{eq:CB_p}
\end{eqnarray}
Therefore, Eq.~\eqref{eq:etaJH_p} becomes:
\begin{eqnarray}
    \eta &=& \nu_e\bigg(\frac{2A''}{7-p}\eta^{2/3}k_*^{p-3}k_c^{7-p}-C_\text{B}\frac{3-p}{5-p}k_c^2\frac{2A''}{5-p}\eta^{2/3}k_*^{p-3}k_c^{5-p}\bigg),\\
    &=& \nu_e\eta^{2/3}k_*^{p-3}k_c^{7-p}\bigg(\frac{2A''}{7-p}-C_\text{B}\frac{2A''(3-p)}{(5-p)^2}\bigg),\\
    &=& (C''_\text{JH}\Delta)^6\bigg(\frac{2A''}{7-p}\eta^{2/3}k_*^{p-3}k_c^{7-p}\bigg)^{1/2}\eta^{2/3}k_*^{p-3}k_c^{7-p}\bigg(\frac{2A''}{7-p}-C_\text{B}\frac{2A''(3-p)}{(5-p)^2}\bigg),\\
    &=&(C''_\text{JH}\Delta)^6\eta k_*^{\frac{3}{2}(p-3)}k_c^{\frac{3}{2}(7-p)}\bigg(\frac{2A''}{7-p}\bigg)^{3/2}\bigg(1-C_\text{B}\frac{(7-p)(3-p)}{(5-p)^2}\bigg)
    \label{eq:eta_JH_p}
\end{eqnarray}
Factoring out $\eta$ and using $k_c=\pi/\Delta$ gives:
\begin{eqnarray}
    C''_\text{JH} = \frac{1}{\pi}\bigg(\frac{k_c}{k_*}\bigg)^{\frac{1}{4}(p-3)}\bigg(\frac{7-p}{2A''}\bigg)^{1/4}\bigg(1-C_\text{B}\frac{(7-p)(3-p)}{(5-p)^2}\bigg)^{-1/6}.\label{eq:CJH_CB_p}
\end{eqnarray}
Note that if $p=3$ and $C_\text{B}=0$, Eq.~\eqref{eq:CJH_CB_p} will be the same as Eq.~\eqref{eq:CHL}. However, if $p=3$, but $C_\text{B}\neq 0$, $C''_\text{JH}$ should take the form in Eq.~\eqref{eq:CHL_CB}.\\

Similarly, $C''_\text{S}$ can be formulated by assuming that $\nu_e$ from Smag and Leith should be the same for a given $\bar{\omega}$ as in Eq.~\eqref{eq:nu_e_being_equal}. The relationship between $C''_\text{S}$ and $C''_\text{L}$ can also be modified by:
\begin{eqnarray}
    \frac{\Delta \langle(\nabla\bar{\omega})^2\rangle^{1/2}}{\langle\bar{\mathcal{S}}^2\rangle^{1/2}} &=& \Delta\frac{\bigg(\frac{2A''}{5-p}\bigg)^{1/2}k_c^{\frac{1}{2}(5-p)}}{\bigg(\frac{2A''}{3-p}\bigg)^{1/2}k_c^{\frac{1}{2}(3-p)}},\\
    &=&\pi\bigg(\frac{3-p}{5-p}\bigg)^{1/2}.\label{eq:ratio_leith_smag_p}
\end{eqnarray}
Following Eq.~\eqref{eq:nu_e_being_equal}, we have:
\begin{eqnarray}
    (C''_\text{S})^2 = (C''_\text{L})^3\bigg(\frac{3-p}{5-p}\bigg)^{1/2} \pi.\label{eq:nu_e_being_equal_p}
\end{eqnarray} 
With Eq.~\eqref{eq:CL_p}, we have:
\begin{eqnarray}
    C''_\text{S}=\frac{1}{\pi}\bigg(\frac{5-p}{2A''}\bigg)^{3/4}\bigg(\frac{3-p}{5-p}\bigg)^{1/4}\bigg(\frac{k_c}{k_*}\bigg)^{\frac{3}{4}(p-3)} .\label{eq:CS_p}
\end{eqnarray} 
Note that if $p=3$, $C''_\text{S}$ should take the form in Eq.~\eqref{eq:CS}. The \(p=4\) case is used only as an effective isotropic representation of the angle-integrated spectrum in Case 3.

\subsubsection{Relationships between $C''_\text{L}$, $C''_\text{S}$, $C''_\text{JH}$, and $C_\text{B}$ and $A''$ for $p = 4$}
Following the derivation in Secs.~\ref{sec:leith}-~\ref{sec:Cs_CL_relation}, for $p = 4$ we have:
\begin{eqnarray}
    \langle\bar{\mathcal{S}}^2\rangle &=& 2\int_0^{k_c} k^2 \hat{E}(k)dk
    \approx 2A''\eta^{2/3}k_*/k_1, \label{eq:S_sq_p_2}\\
     \langle(\nabla \bar{\omega})^2\rangle 
&=& 2\int_0^{k_c} k^4\hat{E}(k)dk =  2A''\eta^{2/3}k_*k_c,\label{eq:grad_vor_sq_p_2}\\
\langle(\nabla^2 \bar{\omega})^2\rangle 
&=& 2\int_0^{k_c} k^6\hat{E}(k)dk = \frac{2A''}{3}\eta^{2/3}k_*k_c^{3}.\label{eq:grad_2_vor_sq_p_2}
\end{eqnarray}

With Eqs.~\eqref{eq:S_sq_p_2}-~\eqref{eq:grad_2_vor_sq_p_2}, we can derive $C''_\text{L}$, $C''_\text{S}$, and $C''_\text{JH}$ (with $C_\text{B}$), accordingly. For example, Eq.~\eqref{eq:eta_leith} gives:
\begin{eqnarray}
\eta=(C''_\text{L}\Delta)^3\bigg(2A''\eta^{2/3}k_*k_c\bigg)^{3/2}=(C''_\text{L}\Delta)^3(2A'')^{3/2}\eta k_*^{\frac{3}{2}}k_c^{\frac{3}{2}}.
\end{eqnarray}
Factoring out $\eta$ and using $k_c=\pi/\Delta$ gives:
\begin{eqnarray}
    C''_\text{L} = \frac{1}{\pi}(2A'')^{-1/2}\bigg(\frac{k_c}{k_*}\bigg)^{\frac{1}{2}}=\frac{1}{\pi}\bigg(\frac{k_c}{2A''k_*}\bigg)^{1/2}.\label{eq:CL_p_2}
\end{eqnarray}
Similarly, Eq.~\eqref{eq:eta_hyper_backscatter} becomes:
\begin{eqnarray}
\eta = \nu_e\frac{2A''}{3}\eta^{2/3}k_*k_c^{3} - \nu_\text{B}2A''\eta^{2/3}k_*k_c. \label{eq:etaJH_p_2}
\end{eqnarray}
Further, similar to Eq.~\eqref{eq:CB}, we have:
\begin{eqnarray}
    \nu_\text{B} &=& -C''_{\text{B}}\frac{\left\langle \bar{\psi}\nu_e \nabla^4\bar{\omega}\right\rangle}{\langle\bar{\psi}\nabla^2\bar{\omega}\rangle} 
= C_\text{B}\nu_e\frac{2\int_0^{k_c} k^2\hat{Z}(k)dk}{2\int_0^{k_c} \hat{Z}(k)dk}= C_\text{B}\nu_e\frac{2\int_0^{k_c} k^4\hat{E}(k)dk}{2\int_0^{k_c}k^2 \hat{E}(k)dk},\\
&=&C_\text{B}\nu_e\frac{2A''\eta^{2/3}k_*k_c}{2A''\eta^{2/3}k_*/k_1} = C_\text{B}\nu_ek_ck_1.
\label{eq:CB_p_2}
\end{eqnarray}
Therefore, Eq.~\eqref{eq:etaJH_p_2} becomes:
\begin{eqnarray}
    \eta &=& \nu_e(\frac{2A''}{3}\eta^{2/3}k_*k_c^{3}-C_\text{B}k_ck_12A''\eta^{2/3}k_*k_c),\\
    &=& 2\nu_e\eta^{2/3}k_*k_c^2A''\bigg(\frac{k_c}{3}-C_\text{B}k_1\bigg),\\
    &=& (C''_\text{JH}\Delta)^6\bigg(\frac{2A''}{3}\eta^{2/3}k_*k_c^3\bigg)^{1/2}2\eta^{2/3}k_*k_c^{2}A''\bigg(\frac{k_c}{3}-C_\text{B}k_1\bigg),\\
    &=&(C''_\text{JH}\Delta)^6\eta k_c^{9/2}\bigg(\frac{2A''k_*}{3}\bigg)^{3/2}\bigg(1-C_\text{B}\frac{3k_1}{k_c}\bigg)
    \label{eq:eta_JH_p_2}
\end{eqnarray}
Factoring out $\eta$ and using $k_c=\pi/\Delta$ gives:
\begin{eqnarray}
    C''_\text{JH} = \frac{1}{\pi}\bigg(\frac{2A''k_*}{3k_c}\bigg)^{-1/4}\bigg(1-C_\text{B}\frac{3k_1}{k_c}\bigg)^{-1/6}.\label{eq:CJH_CB_p_2}
\end{eqnarray}
Note that if $C_\text{B}=0$, Eq.~\eqref{eq:CJH_CB_p_2} denote the coefficient of the hyperviscosity model without backscattering, analogous to Eq.~\eqref{eq:CHL}.\\

Similarly, $C''_\text{S}$ can be formulated by assuming that $\nu_e$ from Smag and Leith should be the same for a given $\bar{\omega}$ as in Eq.~\eqref{eq:nu_e_being_equal}. The relationship between $C''_\text{S}$ and $C''_\text{L}$ can also be modified by:
\begin{eqnarray}
    \frac{\Delta \langle(\nabla\bar{\omega})^2\rangle^{1/2}}{\langle\bar{\mathcal{S}}^2\rangle^{1/2}} &=& \Delta\frac{\bigg(2A''\eta^{2/3}k_*k_c\bigg)^{1/2}}{\bigg(2A''\eta^{2/3}k_*/k_1\bigg)^{1/2}},\\
    &=&\pi(k_1/k_c)^{1/2}.\label{eq:ratio_leith_smag_p_2}
\end{eqnarray}
Following Eq.~\eqref{eq:nu_e_being_equal}, we have:
\begin{eqnarray}
    (C''_\text{S})^2 = (C''_\text{L})^3(k_1/k_c)^{1/2} \pi.\label{eq:nu_e_being_equal_p_2}
\end{eqnarray} 
With Eq.~\eqref{eq:CL_p_2}, we have:
\begin{eqnarray}
    C''_\text{S}=\frac{1}{\pi}\bigg(\frac{k_c^{2/3}k_1^{1/3}}{2A''k_*}\bigg)^{3/4}.\label{eq:CS_p_2}
\end{eqnarray} 

In summary, for a general scaling law $\hat{E}(k)\sim k^{-p}$, $C''_\text{L}$, $C''_\text{S}$, and $C''_\text{JH}$ (with $C_\text{B}$) are functions of $A''$, $k_c$, and $k_*$. Here, $A''$ is the scaling coefficient that can be obtained from the DNS spectrum, and $k_*$ is a characteristic wavenumber. Unlike $k_f$ in Sec~\ref{sec:log_scale}, which is the enstrophy injection wavenumber (forcing wavenumber), $k_*$ does not have a specific physical meaning, and the choice of $k_*$ will not affect the results: the change in $k_*$ will be compensated by the change in $A''$ and $C''_\text{L}$, $C''_\text{S}$, and $C''_\text{JH}$ (with $C_\text{B}$) remain unchanged.\\

\subsubsection{Relationship between general $p$ scaling and log correction scaling}
The general scaling $\hat{E}(k)\sim k^{-p}$ can also be extended to the case of a slowly varying exponent $p(k)$:
\begin{eqnarray}
    \hat{E}(k)=A''\eta^{2/3}k^{-3}(k_*/k)^{p(k)-3},
\end{eqnarray}
where $p(k)$ varies slowly with wavenumber $k$ over the inertial range. If $p(k)$ weakly deviates from $3$, we can write it as:
\begin{eqnarray}
    p(k) = 3+\epsilon(k),~|\epsilon(k)| \ll1.
\end{eqnarray}
If we define $\epsilon(k)$ as:
\begin{eqnarray}
    \epsilon(k) \equiv \frac{qln(ln(k/k_*))}{ln(k/k_*)},
\end{eqnarray}
we have:
\begin{eqnarray}
    \hat{E}(k)&=&A''\eta^{2/3}k^{-3}(k_*/k)^{\epsilon(k)},\\
    &=&A''\eta^{2/3}k^{-3}e^{\frac{qln(ln(k/k_*))}{ln(k/k_*)}ln(k_*/k)},\\
    &=&A''\eta^{2/3}k^{-3}e^{-qln(ln(k/k_*))},\\
    &=&A''\eta^{2/3}k^{-3}[ln(k/k_*)]^{-q}.
\end{eqnarray}

Then, if we use $k_f$ as the characteristic wavenumber $k_*$, and use $q=1/3$ as in~\cite[]{kraichnan1971inertial}, we have the $\hat{E}(k)= A''\eta^{2/3}k^{-3}[ln(k/k_f)]^{-1/3}$ scaling law.\\

\section{Sensitivity analysis of model parameters regarding the fitting range parameter $\xi$}
\begin{figure}
\vspace{.0in}
 \centering
 \vspace*{2mm}
 \begin{overpic}[width=1\linewidth,height=0.3\linewidth]{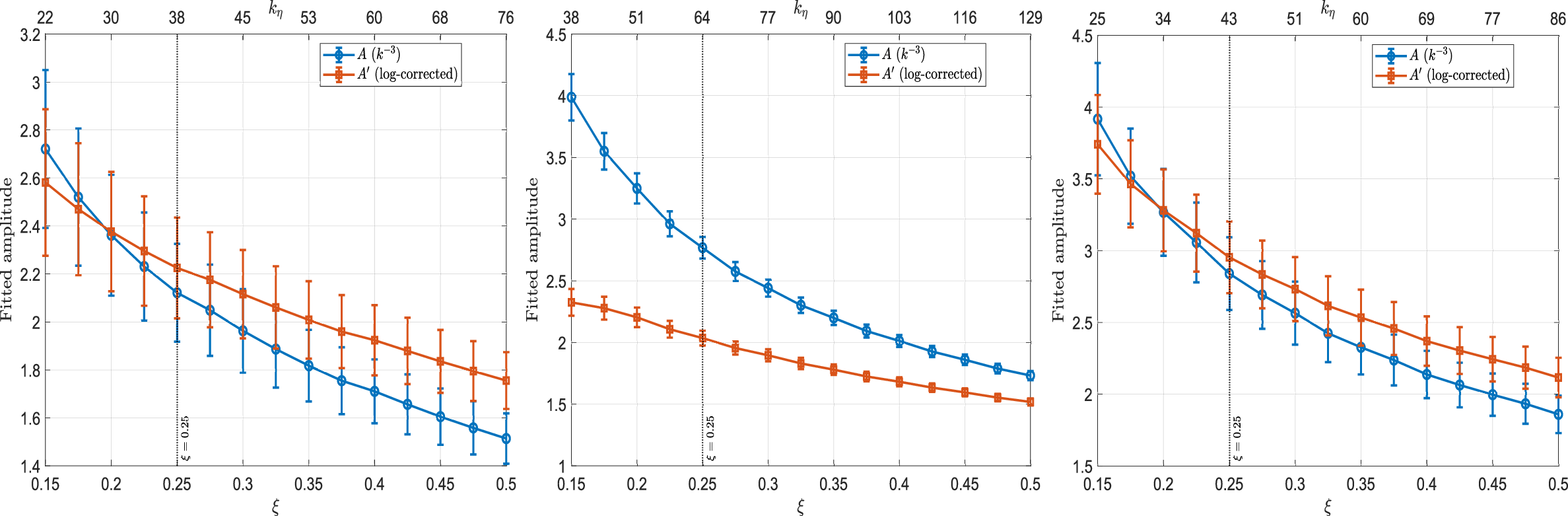}
 \put(11,30.5){{Case 1.1}}
 \put(45,30.5){{Case 2.1}}
 \put(81,30.5){{Case 3}}

 \end{overpic}
  \vspace*{-4mm}
  \caption{\footnotesize Sensitivity of the fitted amplitudes $A$ and $A'$ to the fitting-range prefactor $\xi$. The upper $x$ axis shows the corresponding fitting upper bound $k_\eta$. Error bars show one standard deviation over the 100 DNS snapshots. The log-corrected fit generally shows a weaker dependence on $\xi$ than the pure $k^{-3}$ fit.}
 \label{fig:SI1 A Aprime}
 \vspace*{0mm}
\end{figure}

\begin{figure}
\vspace{.0in}
 \centering
 \vspace*{2mm}
 \begin{overpic}[width=1\linewidth,height=0.4\linewidth]{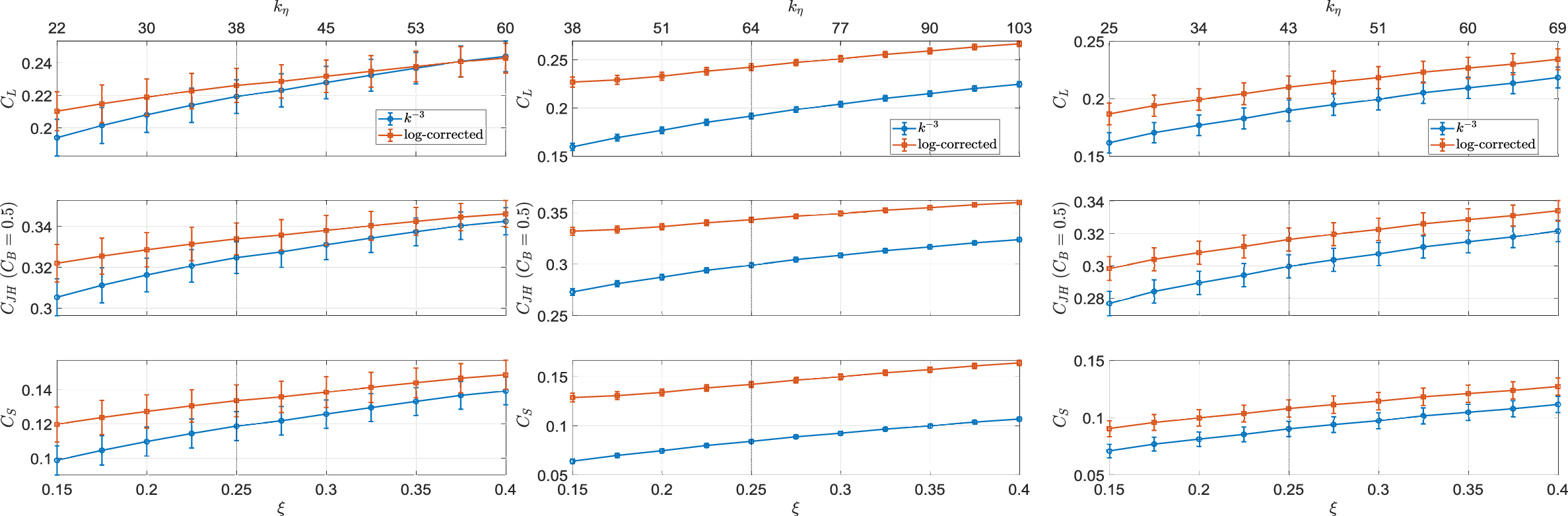}
 \put(11,40.5){{Case 1.1}}
 \put(45,40.5){{Case 2.1}}
 \put(81,40.5){{Case 3}}

 \end{overpic}
  \vspace*{-4mm}
  \caption{\footnotesize Sensitivity of the semi-analytical model parameters $C_\text{L}$, $C_\text{JH}$, and $C_\text{S}$ to the fitting-range prefactor $\xi$. The upper $x$ axis shows the corresponding fitting upper bound $k_\eta$ values with respect to each factor $\xi$. Error bars show one standard deviation over the 100 DNS snapshots. Although the absolute fitted amplitudes depend on the selected fitting range, the induced variation in the closure coefficients remains moderate over the physically reasonable range of $\xi$ considered here. The log-corrected scaling generally reduces the fitting-range dependence.}
 \label{fig:SI2 C Cprime}
 \vspace*{0mm}
\end{figure}

Fig.~\ref{fig:SI1 A Aprime} performs the sensitivity analysis of the scaling law amplitudes $A$ and $A'$ obtained by fitting the scaling laws to the TKE spectra within $k\in[k_f+1,k_\eta]$.  Fig.~\ref{fig:SI2 C Cprime} performs the sensitivity analysis of the model parameters $C_\text{L}$, $C_\text{JH}$, and $C_\text{S}$ obtained by fitting the scaling laws to the TKE spectra within $k\in[k_f+1,k_\eta]$.  The purpose of this sensitivity analysis is to quantify the uncertainty introduced by choosing the upper bound of the spectral fitting range. We do
not interpret $\xi$ as a universal constant. Rather, $\xi$ controls how far the fit extends toward the dissipative roll-off. Smaller $\xi$ uses a more conservative but shorter fitting range, whereas larger $\xi$ includes more wavenumbers but may be increasingly affected by dissipation and bottleneck-like
effects. Therefore, some variation of $A$, $A'$, and the derived closure coefficients with $\xi$ is expected. Despite this expected dependence, the qualitative conclusions are robust over the tested range of $\xi$. The log-corrected scaling generally exhibits weaker fitting-range dependence than the pure $k^{-3}$ scaling, and the resulting $C_\text{L}$, $C_\text{JH}$, and $C_\text{S}$ remain within the same range as the EKI-optimized values reported in the main text. The error bars in Figs. 1-2 denote snapshot-to-snapshot variability for a fixed \(\xi\), while the variation with \(\xi\) represents a systematic fitting-range uncertainty.

\clearpage

\bibliographystyle{my_jfmstyle}
\bibliography{EKI2D}

\end{document}